\documentclass[aps,pre,preprint,onecolumn,showpacs,amsmath,amssymb]{revtex4-1}
\usepackage[T1]{fontenc}
\usepackage{amsmath}
\usepackage{graphicx}
\usepackage{bm}
\usepackage{color}
\usepackage{lineno}
\usepackage{soul}
\usepackage{textgreek}
\usepackage[utf8x]{inputenc}
\usepackage{xcolor}
\usepackage{url}
\usepackage{bm}

\begin{document}

\title{Strain rate controls alignment in growing bacterial monolayers}

\author{Blake Langeslay$^1$}

\author{Gabriel Juarez$^2$}
\thanks{Email address.}\email{gjuarez@illinois.edu}

\affiliation{$^1$Department of Physics, 
University of Illinois Urbana-Champaign, Urbana, Illinois 61801, USA}

\affiliation{$^2$Department of Mechanical Science and Engineering, 
University of Illinois Urbana-Champaign, Urbana, Illinois 61801, USA}

\date{\today}


\begin{abstract}

Growing monolayers of rod-shaped bacteria exhibit local alignment similarly to extensile active nematics. When confined in a channel or growing inward from a ring, the local nematic order of these monolayers changes to a global ordering with cells throughout the monolayer orienting in the same direction. The mechanism behind this phenomenon is so far unclear, as previously proposed mechanisms fail to predict the correct alignment direction in one or more confinement geometries. We present a strain-based model relating net deformation of the growing monolayer to the cell-level deformation resulting from single-cell growth and rotation, producing predictions of cell orientation behavior based on the velocity field in the monolayer. This model correctly predicts the direction of preferential alignment in channel-confined, inward growing, and unconfined colonies. The model also quantitatively predicts orientational order when the velocity field has no net negative strain rate in any direction. We further test our model in simulations of expanding colonies confined to spherical surfaces. Our model and simulations agree that cells away from the origin cell orient radially relative to the colony's center. Additionally, our model's quantitative prediction of the orientational order agrees with the simulation results in the top half of the sphere but fails in the lower half where there is a net negative strain rate. The success of our model bridges the gap between previous works on cell alignment in disparate confinement geometries and provides insight into the underlying physical effects responsible for large-scale alignment.

\end{abstract}

\maketitle

\section{Introduction}

The phenomenon of self organization is of great interest in the fields of biological and active matter \cite{vicsek2,vicsek1}. In animals self organization controls flocking and swarming behaviors \cite{buhl,ward,hayakawa,moussaid}, in epithelial and other cell layers it controls the morphogenesis process \cite{guillamat, saw, endresen, turiv, kawaguchi}, and in bacteria it controls the shape of the colony and the formation and properties of biofilms \cite{copenhagen,yaman,li,xu}. While most cases of self-organization involve motility of the individual organisms, nonmotile bacteria can also exhibit self organization driven only by their extensile growth \cite{dellarciprete,volfson}.

Growing monolayers of nonmotile rod-shaped bacteria can, under certain confinement conditions, self organize to produce large-scale alignment \cite{volfson,you2,isensee,basaran}. In experiments and simulations using either channel-like or inward-growing confinement geometries, the rod-shaped cells that make up the monolayer orient in the same direction across length scales spanning the entire system. Monolayers confined to channels, where growth is more restricted in one dimension than another, produce global alignment parallel to the unconfined axis \cite{volfson,you2,isensee}. Monolayers growing inward, either due to initial cell placement in a circle or confinement to a circular region, result in global radial orientation \cite{basaran}. However, unconfined monolayers whose growth is unrestricted in any direction within the $x-y$ plane do not result in any long-range preferential alignment \cite{you2, you2, dellarciprete}.

Several theories have been proposed to explain this phenomenon. The expanding flow in a channel has been proposed to produce alignment in the direction of alignment \cite{volfson}. However, the unconfined case presents a counterexample to this theory, as its radial expansion does not produce radial alignment \cite{you2}. Anisotropic stress within the monolayer has been claimed to result in cell reorientation toward the lower-stress direction \cite{you2}. However, this mechanism has been shown to break down in highly aligned monolayers, where stress can actually be higher in the direction of alignment \cite{isensee, langeslay2}. In the inward-growing case, the nonlinear velocity profile was cited as the mechanism driving alignment \cite{basaran}. However, in the channel case, the velocity profile is linear and alignment still occurs \cite{isensee, langeslay2}. In summary, no theory of large-scale alignment has been able to predict the presence and direction of preferential alignment across the three simple confinement cases of channels, inward growth, and unconfined growth.

Bacterial monolayers are commonly modeled as composed of hard spherocylindrical particles. Previous work has shown particle-scale effects to be critical to the collective behavior of these colonies, resulting in the formation of highly aligned microdomains \cite{you1, langeslay1}, stress decoupling \cite{isensee}, or complex planar anchoring behavior \cite{langeslay2}. Based on this, we propose a particle-level model of large-scale alignment. The growth and motion of an individual cell produces a local deformation that varies based on its orientation. If the strain rate of the overall monolayer is known, then we can place a constraint on the average cell alignment by hypothesizing that the average local deformation associated with individual cells must equal the net deformation captured in the strain rate.

Applying this method to the previously listed simple confinement cases, we can successfully predict the presence and direction of large-scale alignment in all cases. Additionally, when there is no net negative strain in any direction and the rate of change of cell density is known, we can quantitatively predict the degree of alignment. We further test this method by applying it to the new confinement case of cells growing on the surface of a sphere, where it correctly predicts alignment direction everywhere on the surface and accurately predicts the quantitative orientational order on the upper half of the sphere (but not on the lower half, where there the net strain profile has a negative component).

\section{Results}

\subsection{Strain element model of cell alignment}

To connect the net deformation of a region of the monolayer to the behavior of individual cells, we introduce the concept of a strain element $\boldsymbol{\epsilon}$. This is defined as the strain rate associated with the instantaneous motion of a single cell. If the monolayer is made up entirely of these cells at a constant density, the average strain element $\langle \boldsymbol{\epsilon} \rangle$ over all cells in a region should be equal to the net strain $\boldsymbol{E}$ of the monolayer in that region. 

Let cells be modeled as spherocylinders with diameter 1 \textmu m and length $l$ \textmu m between hemispherical endcaps. Growth is described by the cell growth rate $g$ \textmu m/hr, which gives a linear increase in $l$ over time. When reaching the division length $l_{d}$, a cell divides into two cells with lengths $l_{min}=(l_{d}-1)/2$. Consider a single cell in the basis $(\hat{x'},\hat{y'})$, where $\hat{x'}$ is parallel to the cell's long axis and $\hat{y'}$ is perpendicular while remaining in the plane of the monolayer. The velocity field associated with this cell's motion (neglecting translational velocity, which does not contribute to strain rate) is then
\begin{equation}
v'=\frac{gx}{l+1}\hat{x'}+\dot{\alpha}x\hat{y'}
\end{equation}
where $\alpha$ is the cell's orientation angle with respect to an external basis and $\dot{\alpha}$ is the cell's rotation rate in-plane. This results in a strain rate tensor
\begin{equation}
\boldsymbol{\epsilon}'=\begin{bmatrix}
\frac{g}{l+1} & \dot{\alpha}\\
0 & 0
\end{bmatrix}
\end{equation}
Returning to the external basis $(\hat{x},\hat{y})$ by a rotation through the cell orientation angle $\alpha$ results in
\begin{equation}
\boldsymbol{\epsilon}=\begin{bmatrix}
\frac{g}{l+1}\cos^{2}{\alpha}+\dot{\alpha}\sin{\alpha}\cos{\alpha} & -\frac{g}{l+1}\cos{\theta}\sin{\theta}+\dot{\alpha}cos^{2}{\alpha}\\
\frac{g}{l+1}\cos{\theta}\sin{\theta}-\dot{\alpha}sin^{2}{\alpha} & \frac{g}{l+1}\sin^{2}{\alpha}-\dot{\alpha}\sin{\alpha}\cos{\alpha}
\end{bmatrix}
\end{equation}

If we choose the basis so that $\hat{x}$ is the direction of colony expansion (normal to its boundary) and assume that cells are equally likely to orient tilted away from the boundary in either direction, the distributions of $\alpha$ and $\dot{\alpha}$ will be symmetric. The off-diagonal elements of $\boldsymbol{\epsilon}$ are therefore antisymmetric and vanish when averaging over many cells. The resulting tensor can be written as follows:
\begin{equation}
\boldsymbol{\epsilon}=\frac{g}{l+1}\begin{bmatrix}
\cos^{2}{\alpha}& 0\\
0 & \sin^{2}{\alpha}
\end{bmatrix}+\dot{\alpha}\sin{\alpha}\cos{\alpha}\begin{bmatrix}
1 & 0\\
0 & -1
\end{bmatrix}=\boldsymbol{G}+\boldsymbol{A}
\end{equation}
where $\boldsymbol{G}$ and $\boldsymbol{A}$ are the parts of the strain rate sourced from growth and rotation, respectively. The growth strain $\boldsymbol{G}$ is always positive and higher in the direction of cell orientation. The alignment strain $\boldsymbol{A}$ always has one positive and one negative component. If the cell is rotating toward the direction of colony expansion then $A_{xx}$ is positive and $A_{yy}$ is negative, and if the cell is rotating toward $\hat{y}$ the reverse is true.

Combining this with the assumption that the average strain element is equal to the net strain allows qualitative prediction of cell orientation direction. On average, cells must orient or be rotating toward the direction of the highest extensional strain. Additionally, if negative strain is present, cells must be rotating away from the direction of most negative strain.

\subsubsection{Testing strain element prediction in sample geometries}
\begin{figure*}
\centering
	\includegraphics[width=0.7\linewidth]{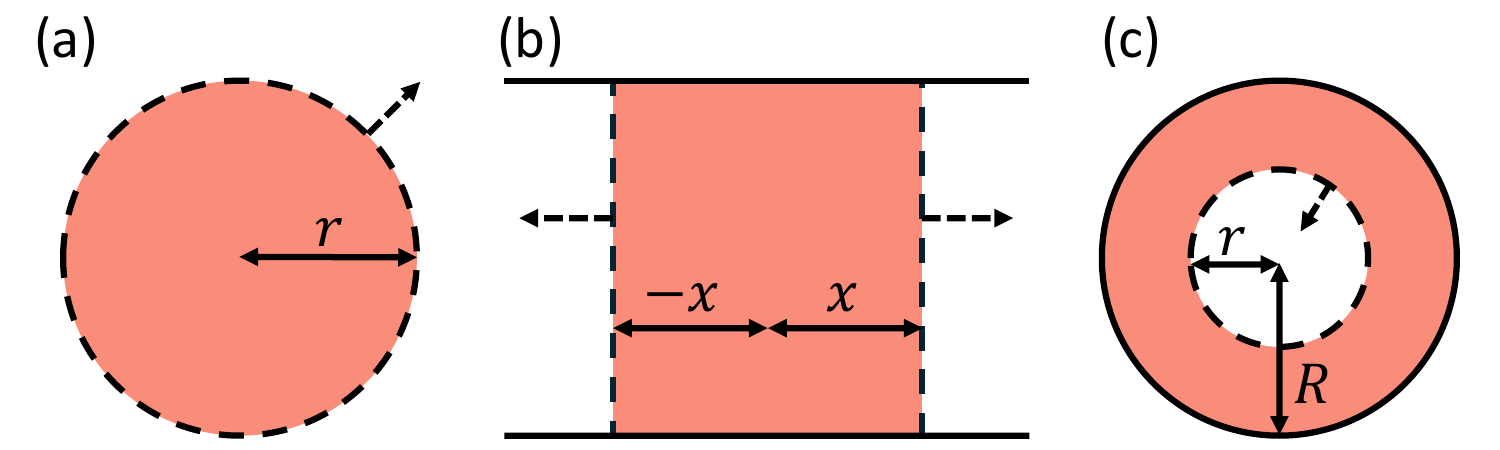}
\caption{
Schematics of differently confined growing bacterial colonies. Dotted boundaries represent the expanding colony edge, dotted arrows represent the direction of expansion. (a) Unconfined growing monolayer. The monolayer forms a circle with radius $r$ increasing as it expands. (b) Growing monolayer in a channel with periodic boundaries (solid horizontal lines). Colony edges are at $\pm x$. (c) Inward-growing monolayer, confined within a circular outer wall (solid outer line) at radius $R$. The monolayer forms a ring with inner radius $r$ decreasing as it expands. 
}
	\label{fig:figone}
\end{figure*}

To test this prediction, we choose three simple confinement geometries where alignment behavior is already known: no confinement (\ref{fig:figone}(a)), confinement in a channel with periodic boundaries in the $y$ direction and outlets (boundaries past which cells are removed from the experiment) in the $x$ direction (\ref{fig:figone}(b)), and inward growth confined by a circular outer wall of radius $R$ (\ref{fig:figone}(c)). We assume that the monolayer has the same symmetry as the confinement geometry and that its density is constant. The monolayer's expansion is governed solely by the exponential growth constant $k_{g}$:
\begin{equation}
\label{eqn:growth}
\frac{dA}{dt}=k_{g}A
\end{equation}
where $A$ is the area of any portion of the monolayer. Applying this to the confinement geometries listed leads to the following velocity profiles (derivations in Appendix I):
\begin{equation}
v_{r}=k_{g}r/2
\end{equation}
\begin{equation}
v_{x}=k_{g}x
\end{equation}
\begin{equation}
v_{r}=-k_{g}(R^{2}-r^{2})/2r
\end{equation}
for unconfined, channel, and inward growth, respectively. The other velocity components ($v_{\theta}$ for unconfined and inward growth, $v_{y}$ for channel growth) are zero.

Beginning with the unconfined case in the basis $(\hat{r},\hat{\theta})$, we obtain the net strain tensor $\boldsymbol{E}=\frac{k_{g}}{2}\boldsymbol{I}$ (derivation in Appendix II). In this case, strain rate is isotropic and there should be no direction of preferential alignment. This is precisely what has been found in unconfined experiments and simulations \cite{dellarciprete, you1}.

In the channel case, the only nonzero component of the net strain tensor is $E_{xx}=k_{g}$. This correctly predicts that cells preferentially align to the $x$-axis, along the length of the channel \cite{volfson}. Additionally, the presence of outlets in this geometry allows the system to reach a steady state in which $\boldsymbol{A}=0$ and all strain must be sourced from the cell growth term $\boldsymbol{G}$. In this case, cells must actually align perfectly to the $x$-axis so that $\epsilon_{yy}$ vanishes. This again matches previous results for the steady-state channel case \cite{orozcofuentes, you2, isensee}. 

In the inward growth case (again in the basis $(\hat{r},\hat{\theta})$), the strain tensor is as follows:
\begin{equation}
\boldsymbol{E}=\frac{k_{g}}{2r^{2}}\begin{bmatrix}
r^{2}+R^{2} & 0\\
0 & r^{2}-R^{2}
\end{bmatrix}
\end{equation}
For $r<R$, $E_{rr}$ is always positive and $E_{\theta \theta}$ is always negative, so cells align radially in accordance with previous results \cite{basaran}. 

\subsubsection{Predicting orientational order}
If we assume that $\boldsymbol{G}>>\boldsymbol{A}$, as in the steady-state channel case above, then the strain element model can be used to quantitatively predict the orientational order of cells in the monolayer. The orientational order $S_{i}$ describes the degree of alignment of a cell relative to the unit vector $\hat{i}$:
\begin{equation}
\label{eqn:order}
S_{i}=\cos^{2}{\alpha}-1
\end{equation}
where $\alpha$ is the angle between the cell orientation vector and $\hat{i}$. Here $\langle S_{i} \rangle=1$ describes perfect alignment in the $\hat{i}$ direction, $\langle S_{i} \rangle=-1$ describes perfect alignment perpendicular to $\hat{i}$, and $\langle S_{i} \rangle=0$ describes no preferential alignment. 

We can write the average growth portion $\langle \boldsymbol{G} \rangle$ of the strain element in terms of $\langle S_{i} \rangle$:
\begin{equation}
\langle \boldsymbol{G} \rangle=\biggl \langle \frac{g}{l+1}\begin{bmatrix}
\cos^{2}{\alpha}& 0\\
0 & \sin^{2}{\alpha} 
\end{bmatrix} \biggr \rangle
\end{equation}
Assuming that $g$, $l$, and $\alpha$ are independent, this gives
\begin{equation}
\label{eqn:a}
\langle \boldsymbol{G} \rangle= \langle g \rangle \biggl \langle \frac{1}{l+1} \biggr \rangle \begin{bmatrix}
\langle \cos^{2}{\alpha} \rangle& 0\\
0 & \langle \sin^{2}{\alpha} \rangle 
\end{bmatrix}=
g_{0} \biggl \langle \frac{1}{l+1} \biggr \rangle \begin{bmatrix}
(1+\langle S_{i} \rangle)/2& 0\\
0 & (1-\langle S_{i} \rangle)/2 
\end{bmatrix}
\end{equation}
where $g_0$ is the average growth rate. Defining the cells' division length $l_{d}$ as the value of $l$ at which they divide, we can simplify eqn. \ref{eqn:a} in terms of $k_{g}$ (derivation in Appendix III):
\begin{equation}
\label{eqn:b}
\langle \boldsymbol{G} \rangle=
k_{g} \begin{bmatrix}
(1+\langle S_{i} \rangle)/2& 0\\
0 & (1-\langle S_{i} \rangle)/2 
\end{bmatrix}
\end{equation}

Using our assumption that the average strain element is equal to the net strain (and the further assumption that we can neglect $\boldsymbol{A}$), we obtain
\begin{equation}
\label{eqn:order2}
\langle S_{i} \rangle=2E_{ii}/k_{g}-1
\end{equation}
In cases where $E_{ii}>k_{g}$ or $E_{ii}<0$, such as the inward growth case described above, this leads to an inconsistency as $S$ cannot be greater than 1 or less than -1. In these cases, one or more of the basic assumptions of the model must break down (see Discussion for more details). However, in cases where  $0\le E_{ii}\le k_{g}$ (including unconfined and channel growth), this model produces quantitative predictions of orientational order within the monolayer.

Channel growth represents an extreme end of this spectrum, where $E_{xx}=k_{g}$ and $E_{yy}=0$, implying $\langle S_{x} \rangle=1$ and $\langle S_{y} \rangle=0$. This results in perfect alignment to the channel's length, as previously predicted. Unconfined growth represents the other end, in which $E_{rr}=E_{\theta \theta}=k_{g}/2$ and $\langle S_{r} \rangle=\langle S_{\theta} \rangle=0$. This results in a colony with no preferential alignment, again as previously predicted.

\subsection{Simulations of growing monolayers on spherical surfaces}

Expanding colonies with non-equal and nonzero values of both primary strain components are so far unexplored. To test our model's predictions in this range of parameter space we simulated a growing colony on a spherical surface, initialized with a single cell at the north pole ($\theta=0$). Cells near the north pole should approximate the flat unconfined case, as their local neighborhood is approximately flat. Cells near the equator ($\theta=\pi/2$) should approximate the channel confined case, as the circumference of the sphere effectively creates a periodic boundary. Cells near the south pole ($\theta=\pi$) should approximate the inward growth case, as the colony converges on the sphere's farthest point. Intermediate points on the sphere should then produce intermediate combinations of strain rate components.

\begin{figure*}
\centering
	\includegraphics[width=\linewidth]{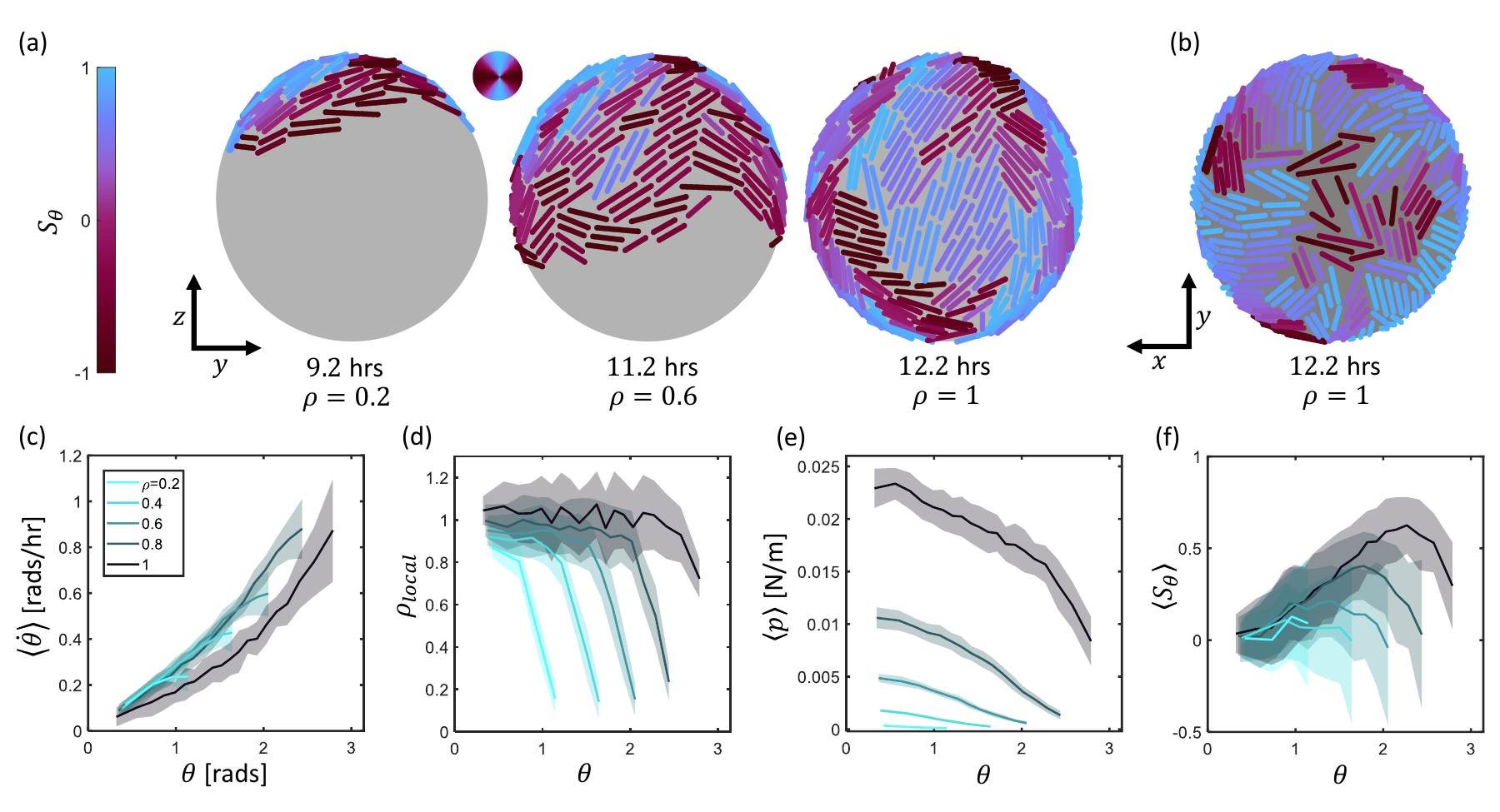}
\caption{
(a) Simulation frames of cells on a spherical substrate viewed from the $+x$ direction. The spherical surface has radius $R=14$ \textmu m.
Cells are plotted at half true diameter for clarity and are colored according to the order parameter $S_{\theta}$ so that light blue cells align to $\hat{\theta}$ and dark red cells align to $\hat{\phi}$. 
Frames from left to right are at surface coverage $\rho=0.2$, $0.6$, $1$ and show the monolayer's downward growth to cover the entire spherical surface. 
(b) Simulation frame at $\rho=1$ showing the sphere's south pole from the $-z$ direction, demonstrating aligned cells converging in an aster-like defect. 
(c) Profiles of average $\dot{\theta}$ at varying surface coverages. 
(d) Profiles of local packing fraction $\rho_{local}$. 
(d) Profiles of average pressure. 
(f) Profiles of average orientational order. 
(c-f) Lighter profiles correspond to lower surface coverage $\rho$. 
All shaded regions represent standard deviation over 48 trials.
}
	\label{fig:figtwo}
\end{figure*}

For the simulations, we used a molecular dynamics model of cell growth and interaction described in the methods. The cell width $d_{0}$ was set to 1 \textmu m, the division length $l_{d}$ to 6 \textmu m, and the radius of the spherical surface to 14 \textmu m. A set of 48 trials was conducted with this setup. In all trials, the growing colony forms a spherical cap that expands downward toward $\theta=\pi$ until the sphere is entirely covered (Fig. \ref{fig:figtwo}a). At longer times, the cells appear to orient toward the $\hat{\theta}$ direction. At the south pole, these aligned cells converge in an aster-like $+1$ charge topological defect (Fig. \ref{fig:figtwo}b).

To quantify how monolayer structure varies over the surface of the sphere, the cell velocity, packing fraction, pressure, and orientational order were averaged over bins in both time and space. The spherical surface was divided by $z$-coordinate into 20 slices of equal surface area. Parameters were then averaged in each bin for a series of time steps with total surface packing fractions of $\rho= 0.2$, 0.4, 0.6, 0.8, and 1.

The motion of the growing monolayer was characterized with $\dot{\theta}$, the component of velocity toward $\hat{\theta}$ normalized by the sphere radius $R$ (Fig. \ref{fig:figtwo}c). The average value of this velocity is always positive (away from the colony center), and increases in magnitude with increasing distance from the center ($\theta=0$). Profiles of $\langle \dot{\theta}\rangle$ show little variation at different surface coverage fractions.

The density of the monolayer was characterized by the local packing fraction $\rho_{local}$ within a slice. The packing fraction is roughly constant and close to 1 within the bulk of the colony, rapidly falling to zero at the colony's edge (Fig. \ref{fig:figtwo}d). Similarly to $\langle \dot{\theta}\rangle$, profiles of $\rho_{local}$ show little variation with increasing surface coverage, although the bulk packing fraction slightly increases over time.

To characterize the forces on cells within the monolayer, the pressure was calculated. 
This was defined based on the Virial stress $\boldsymbol{\sigma}_{i}$ for each cell $i$ \cite{you1, volfson}:
\begin{equation}
    \boldsymbol{\sigma}_{i}=\frac{\phi}{a_{i}}\sum_{j} \boldsymbol{r}_{ij}\boldsymbol{F}_{ij}
\end{equation}
where \(\boldsymbol{r}_{ij}\) is the vector from the center of cell \(i\) to the point of contact with cell \(j\) and \(\boldsymbol{F}_{ij}\) is the force from cell \(j\) on cell \(i\). Based on this stress tensor, the pressure $p$ is as follows \cite{you1}:
\begin{equation}
p=\frac{1}{2}(\sigma_{\theta \theta}+\sigma_{\phi \phi})
\end{equation}
Similar to simulations on flat surfaces \cite{you1}, we find that $\langle p \rangle$ is highest near the colony's center ($\theta=0$) and decreases toward zero at its edge (Fig. \ref{fig:figtwo}e). Pressure also dramatically increases with increasing colony size, although the distribution's form remains similar.

To characterize cell alignment on the sphere, the average orientational order $\langle S_{\theta} \rangle$ was calculated (eqn. \ref{eqn:order}). Simulation snapshots with individual cells colored by $S_{\theta}$ are shown in Fig. \ref{fig:figtwo}(a). Profiles of $\langle S_{\theta} \rangle$ over the surface show that orientation trends toward $\hat{\theta}$ with increasing distance from the colony's center (Fig. \ref{fig:figtwo}f). At the leading edge of the colony's expansion this trend reverses, with $\langle S_{\theta} \rangle$ rapidly decreasing. This edge behavior appears similar to an effect seen in colonies growing on flat surfaces, where cells at the leading edge preferentially align tangent to the edge \cite{you1, dellarciprete, basaran}. Apart from the different positions of the leading edge, profiles of $\langle S_{\theta}\rangle$ at different surface coverages are very similar, with the curves for $\rho=0.8$ and $\rho=1$ nearly coinciding in the upper half of the sphere.

\subsubsection{Strain element predictions for spherical surfaces}

\begin{figure*}
\centering
	\includegraphics[width=8.5cm]{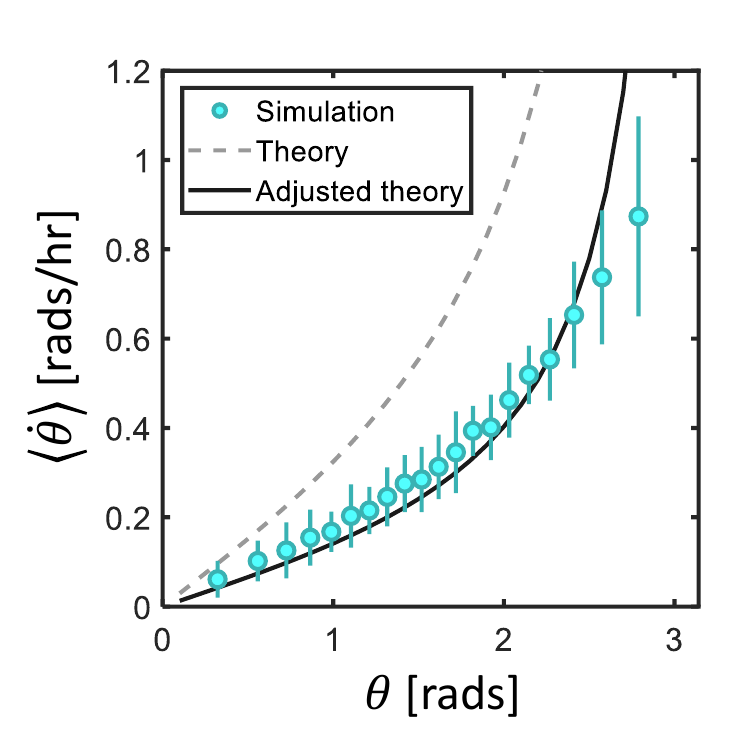}
\caption{
Comparison of $\langle \dot{\theta} \rangle$ data from simulations to analytically predicted profiles (dashed and solid lines). 
Data taken from frames where total surface coverage $\rho=1$. 
Error bars are standard deviation.
}
	\label{fig:figthree}
\end{figure*}

To compare this alignment behavior to the strain element predictions, we first construct the theoretical velocity profile of the expanding colony. Using the same method as for the unconfined, channel, and inward growth cases gives the following (derivation in Appendix I)
\begin{equation}
\label{eqn:vel}
\frac{d\theta}{dt}=
k_{g}\frac{1-\cos{\theta}}{\sin{\theta}}
\end{equation}
The simulation data for $\dot{\theta}$ when total surface coverage $\rho=1$ is compared to this predicted profile (dashed line) (Fig. \ref{fig:figthree}). While the qualitative form of the simulation data matches the theoretically predicted velocity profile, there is a large quantitative error between the two.

This error can be eliminated by relaxing the assumption of constant cell density. It can be seen in Fig. \ref{fig:figtwo}(d) that, while local packing fraction $\rho_{local}$ has minimal spatial variation within the bulk of the colony, it does slightly increase in time. Allowing packing fraction to vary in time leads to a modified form of eqn. \ref{eqn:growth} (derivation in Appendix IV):
\begin{equation}
\frac{dA}{dt}=\left( 1-\frac{1}{k_{g}\rho}\frac{d\rho}{dt}\right) k_{g}A
\end{equation}
which in turn introduces a correction to the velocity profile:
\begin{equation}
\label{eqn:veladjusted}
\frac{d\theta}{dt}=\left( 1-\frac{1}{k_{g}\rho}\frac{d\rho}{dt}\right)k_{g}\frac{1-\cos{\theta}}{\sin{\theta}}
\end{equation}
The packing fraction and its derivative can be found empirically from the simulation data. To avoid the non-uniform packing fraction near the colony's edge, these quantities are calculated in only the top half of the sphere. The resulting adjusted velocity profile is compared to simulation data in Fig. \ref{fig:figthree}. The profiles quantitatively match except at the colony's leading edge where the assumption of spatially invariant packing fraction is already known to fail (Fig. \ref{fig:figtwo}d).

Using this adjusted velocity profile, the net strain tensor in the $\hat{\theta},\hat{\phi}$ basis is (derivation in Appendix II)
\begin{equation}
\boldsymbol{E}=\dot{\theta}\begin{bmatrix}
1/\sin{\theta} & 0\\
0 & 1/\tan{\theta}
\end{bmatrix}
\end{equation}
At $\theta=0$, the strain components $E_{\theta \theta}$ and $E_{\phi \phi}$ are equal. With increasing $\theta$, $E_{\theta \theta}$ increases and $E_{\phi \phi}$ decreases, with $E_{\phi \phi}$ becoming negative at $\theta>\pi/2$. 

The alignment behavior of cells on the sphere can be qualitatively predicted by the strain rate in the same way as the simpler confinement examples previously discussed. Near $\theta=0$, where the strain components are equal, there is no preferential alignment and $S_{\theta}=0$ (Fig. \ref{fig:figtwo}f). Elsewhere, $E_{\theta \theta}>E_{\phi \phi}$, so cells preferentially align to the $\hat{\theta}$ direction and $S_{\theta}>0$. As the difference between the net strain components increases with higher $\theta$, the degree of orientational order also increases.

To directly apply the strain element model here, eqn. \ref{eqn:order2} must be adjusted in the same way as the velocity profile to account for non-constant density. This is accomplished by adding an additional strain term $\textbf{D}$, the density strain, to the relation between the net strain and the average strain element:
\begin{equation}
\label{eqn:strainbalance}
\textbf{E}=\langle \boldsymbol{\epsilon} \rangle+\textbf{D}=\langle \boldsymbol{\epsilon} \rangle-\frac{\dot{\rho}}{2\rho}\textbf{I}
\end{equation}
This accounts for the deformation associated with cells moving closer together (farther apart) as packing fraction increases (decreases).

\begin{figure*}
\centering
	\includegraphics[width=\linewidth]{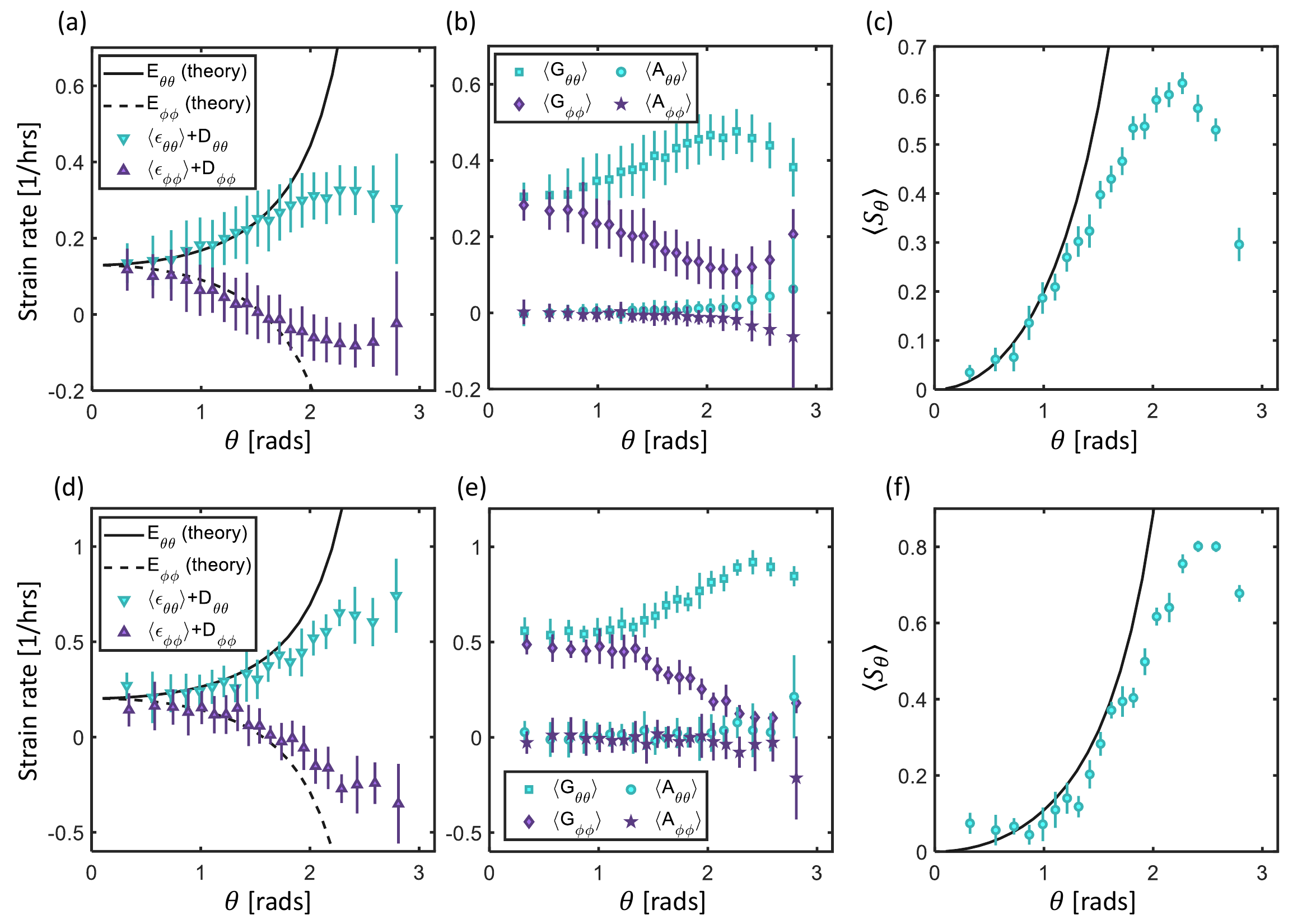}
\caption{
(a) Comparison of average strain elements $\langle \epsilon_{ii} \rangle$ from simulations to theoretical net strain rates $E_{ii}$ (solid and dashed lines) for cells with $l_{d}=6$ \textmu m. Error bars are standard deviation. (b) Comparison of contributions from the growth and alignment parts of the strain element $\textbf{G}$ and $\textbf{A}$, respectively. Error bars are standard deviation. (c) Comparison of average orientational order profile from simulations to analytical prediction (solid line). Error bars are standard error. (d-f) The same plots as (a-c) for cells with $l_{d}=3$.
}
	\label{fig:figfour}
\end{figure*}

The theoretical net strain (using the adjusted velocity from eqn. \ref{eqn:veladjusted}) is compared to the average strain element and density strain from simulations in Fig. \ref{fig:figfour}(a), again for the case where $\rho=1$. In the upper half of the sphere ($\theta>\pi$), eqn. \ref{eqn:strainbalance} quantitatively holds. In the lower half of the sphere, it becomes increasingly inaccurate as the theoretical strain diverges and the simulation data remains finite.

Comparing the growth and alignment parts of the strain element ($\textbf{G}$ and $\textbf{A}$) reveals that the strain element's behavior is dominated by the growth term, especially in the upper half of the sphere (Fig. \ref{fig:figfour}(b)). Only in the lower half of the sphere, where net strain has a negative component, does $\textbf{A}$ increase to magnitudes comparable to $\textbf{G}$. For regions where the net strain has no negative components, the dominance of $\textbf{G}$ over $\textbf{A}$ validates the earlier assumption that $\textbf{A}$ can be neglected when quantitatively predicting orientational order.

To quantitatively predict $\langle S_{\theta} \rangle$ in our simulations, eqn. \ref{eqn:order2} must again be modified to include the density strain $\textbf{D}$. This results in the following:
\begin{equation}
\langle S_{\theta} \rangle=2(E_{ii}-\frac{\dot{\rho}}{2\rho})/k_{g}-1
\end{equation}
Where again $E_{ii}$ is calculated with the adjusted velocity. This can be rewritten in terms of the dimensionless parameter $\gamma=\frac{\dot{\rho}}{2\rho}/k_{g}$:
\begin{equation}
\label{eqn:final}
\langle S_{\theta} \rangle=2(E_{ii}/k_{g}-\gamma)-1
\end{equation}
The predicted orientational order based on this model is compared to simulation data in Fig. \ref{fig:figfour}(c). The model is a good match for $\theta>\pi/2$. For $\theta>\pi/2$ the model is not a good fit, predicting a nonphysical orientational order of $S>1$ because a component of the net strain ($E_{\phi \phi}$) is negative. This effect is explained in more detail in the Discussion.

To test the generality of the strain element model, a set of 12 simulation trials was run with $l_{d}=3$ \textmu m rather than $6$, keeping all other parameters including the average cell growth rate $g_{0}$ identical. The theoretical strain $\boldsymbol{E}$ is compared to the simulation data for $\langle \boldsymbol{\epsilon} \rangle$ in Fig. Fig. \ref{fig:figfour}(d). The theory again matches the data for $\theta<\pi/2$ and diverges past that point, although the divergence is slower than for $l_{d}=6$ (Fig. \ref{fig:figfour}(a)). Plotting the individual strain sources $\boldsymbol{G}$ and $\boldsymbol{A}$ reveals very similar behavior to the $l_{d}=6$ 
\textmu m case, with $\boldsymbol{A}$ nearly zero until near the south pole and $\boldsymbol{G}$ contributing the bulk of the variation in the total strain element (Fig. \ref{fig:figfour}(e)).

The predicted profile of $\langle S_{\theta} \rangle$ for the $l_{d}=3$ simulations again matches the simulation data (Fig. \ref{fig:figfour}(f)). In comparison to the $l_{d}=6$ case (Fig. \ref{fig:figfour}(c)), the lower aspect ratio cells deviate more from the predicted profile at lower $\theta$, with the predicted profile slightly overestimating the actual orientational order. Both the predicted and measured profiles of orientational order are very similar between Figs. 4(c) and 4(f), despite the difference in aspect ratio. This behavior differs significantly from previous studies of bacterial alignment, where increasing cell aspect ratio significantly increases orientational order \cite{you1, isensee, langeslay1}. We can confirm from eqn. \ref{eqn:final} that the predicted profile depends only on $\gamma$. In this specific case, decreasing the division length without changing the growth rate increases the growth constant $k_{g}$ from 0.59 to 1.04 1/hrs, and the value of $\dot{\rho}/\rho$ obtained from simulations simultaneously increases from 0.34 to 0.63 1/hrs. The change in $\dot{\rho}/\rho$ can be intuitively explained as a more rapid increase in surface density in response to the increased growth rate. The combination of these changes leads to a very similar value of $\gamma$ between the two different aspect ratios, and therefore similar profiles of orientational order.

\section{Discussion}

By connecting particle-level strain elements to the net strain in a growing bacterial monolayer, we have created a model able to qualitatively predict the direction of alignment in unconfined, channel confined, and inward growing systems. This model's predictions align with previous experimental and simulation results in equivalent systems \cite{dellarciprete, you1, you2, volfson, isensee, basaran}. Multiple models of the mechanism driving large-scale alignment have previously been suggested; however, no previous model has been successful in predicting alignment across all three of these systems. 

In physical terms, our model states that the net deformation of the monolayer is driven by particle-scale deformations. These can be sourced from either growth or rotation of individual cells, although in simulations we find that the growth term usually dominates. Because of this, the direction of highest extensional deformation must be the direction of preferential cell alignment, as cell growth produces particle-scale extensional deformation in the direction of alignment.

The strain element model also allows quantitative prediction of the orientational order within the monolayer. In practice, accurate predictions require empirical measurement of the packing fraction and its derivative, as changes in cell density contribute to the net deformation. Additionally, the model's calculation of orientational order is only valid in regions where there is no compressive deformation (no diagonal component of the strain tensor is negative), as compression produces definitionally impossible predictions of order greater than 1 (more than perfect alignment). However, despite its limitations, the quantitative accuracy of this calculation in our simulations strongly supports the underlying argument that cell alignment is controlled by net monolayer deformation.

Our model is sufficient to predict alignment behavior in systems with enough symmetry that position can be described with a single variable. This is true in all the example cases we have chosen: in the unconfined and inward growth cases, the system is invariant under rotation through $\theta$, and $r$ completely describes position. In the channel case (with a periodic boundary), the system is invariant under translation in $y$, and $x$ describes position. Similarly, the spherical case is invariant under rotation through $\phi$, and $\theta$ describes position. With less symmetry (for example, in a region bounded by differently-sized outlets \cite{isensee}, or in a channel with an obstruction \cite{langeslay2}) the simple arguments we have used to predict the velocity field (and by extension the strain tensor) are no longer sufficient. In these more complex cases, a hydrodynamic continuum model relating pressure, density, and velocity within the monolayer would be required in order to construct a velocity field and utilize the strain element model to predict alignment.

\begin{figure*}
\centering
	\includegraphics[width=8.5cm]{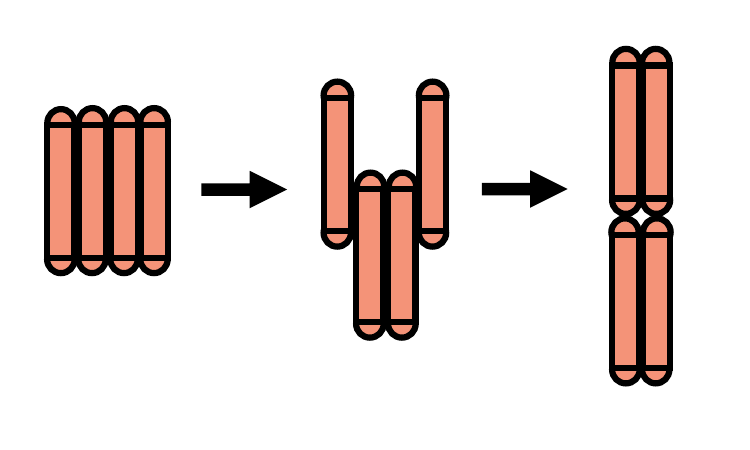}
\caption{
Schematic showing an example of rearrangement strain. The group of cells compresses in the horizontal direction and elongates in the vertical direction without any elongation or rotation of the individual cells. 
}
	\label{fig:figfive}
\end{figure*}

In situations where a component of the net strain is negative, the strain element model breaks down. The only way to generate negative particle-level strain within the model is by rotating cells toward the direction of alignment (within the $\textbf{A}$ term), and this obviously becomes unsustainable as the system approaches perfect alignment. This can be seen in Fig. \ref{fig:figfour}(a,d), where the mean strain element falls far short of the net strain for $\theta>\pi/2$, the region of the sphere where $E_{\phi \phi}$ is negative. The same issue causes the theoretical strain profiles in Fig. \ref{fig:figfour}(c,f) to fail for $\theta>\pi/2$, predicting a nonphysical orientational order of $S>1$. In the case of negative net strain there is evidently another source of deformation not accounted for by the strain element or the density strain. 

We suggest that this deformation could be sourced from rearrangements involving multiple cells, which would not be captured by the single-cell strain element. An example of a rearrangement resulting in net deformation without growth or rotation of individual cells is shown in Fig. \ref{fig:figfive}. As this type of deformation requires significant rearrangement of cells that would be opposed by the liquid crystalline order of the surrounding monolayer, it is unsurprising that their effects are negligible when net strain is strictly positive and can be accounted for by cell growth alone. However, in negative-strain regions where there is no other way for the net strain to be accommodated, the prevalence of these rearrangements could increase.

It is interesting that the orientational order profiles predicted by eqn. \ref{eqn:final} are independent of cell aspect ratio. Aspect ratio has commonly been identified as a controlling factor of orientational order, with higher aspect ratio cells exhibiting stronger alignment. However, the net strain rates $E_{ii}$ are independent of aspect ratio, as are the growth strain rates $G_{ii}$, so this independence is consistent with our physical description of the alignment mechanism. In practice, we find that orientational order is somewhat lower than our model predicts for lower aspect ratio cells (Fig. \ref{fig:figfour}(f)). We argue that this is again due to a breakdown in the assumption that growth and packing fraction change account for all of the net strain. For lower aspect ratio cells, deviation from the liquid crystal alignment of the monolayer is easier. Therefore, rearrangement-type deformations are more possible, leading to non-negligible strain contributions due to rearrangements even in the upper half of the sphere. In this sense our model can be taken as a high aspect ratio limit, although even for the lower aspect ratio case $l_{d}=3$ it appears to hold relatively well.

While the strain element model identifies net monolayer deformation as the controlling factor of large-scale cell alignment, it does not identify a direct physical mechanism responsible for reorientation of individual cells. We suggest that, as originally suggested by Volfson et al., this reorientation is caused by the colony's flow field \cite{volfson}. More specifically, it is well known that elongated particles (including rod-shaped bacteria) in extensional flow align to the principal axis of extension \cite{graham}. Previously the unconfined case has been cited as a counterexample to this mechanism, as its radially expanding flow does not result in any large-scale radial alignment \cite{you2}. However, as shown in the strain rates we calculated for the unconfined case ($E_{rr}=E_{\theta \theta}$), the expanding flow in the unconfined case has no principal direction of extension and therefore no way to generate a preferential alignment direction. Our model holds that alignment always occurs in the direction of highest strain which coincides with the principal axis of extension, so in the absence of a counterexample, extensional flow is an intuitive candidate for a physical mechanism.

More specifically to our simulations on spherical surfaces, we have shown that spherical confinement can produce large-scale alignment in expanding bacterial monolayers. This depends on the same physical mechanisms as previous confinement geometries that have produced large-scacle alignment. However, because the geometric properties of the sphere change with distance from the colony's center at $\theta=0$, its alignment behavior varies throughout its surface. There is no preferential alignment at the colony center, and the degree of alignment increases with increasing $\theta$.

Our results on spherical surfaces have interesting consequences for the case of bacteria growing on droplets. In particular, the case of \textit{A. borkumensis}, an oil-degrading marine bacterium known to grow on and deform droplets of crude oil spilled in the ocean, is of practical importance \cite{hickl1, hickl2, prasad}. A straightforward prediction of our simulations is that when growth on a sphere starts from a single cell, a $+1$ charge aster defect eventually forms opposite the starting position. Based on deformation mechanics of active nematics confined to deformable spheres \cite{keber, metselaar, maroudassacks}, as well as experimental results on bacteria-covered oil droplets in the lab \cite{prasad}, this type of defect is likely to nucleate a tube-like deformation. From previous simulations taking into account cell concentration and droplet size, the case where a droplet encounters only a single cell is likely common \cite{hickl1}. In these cases, we might expect an isolated tube deformation to appear at the droplet's south pole before more widespread buckling of the droplet surface.

In conclusion, we find that alignment in growing bacterial monolayers is well described using a model that connects net strain rate to individual cell behavior (strain elements). This model qualitatively predicts the correct alignment direction for unconfined, channel, and inward growth, where previously no single model was able to consistently explain the direction of preferential growth in these different geometries. In cases where there is no net compressive deformation of the monolayer (no negative net strain components) and the rate of change of its density is known, our model also produces accurate quantitative predictions of orientational order. Applying this model to the new case of a growing colony confined to a spherical surface is successful in the same ways, and reveals several interesting new insights about the behavior of growing bacterial monolayers on curved surfaces. Our results unify several previous avenues of research into growing bacterial monolayers as active matter and are likely to be useful for understanding of growth-driven alignment in future work.

\section{Methods}

\subsection{Molecular Dynamics Model}

Cells were modeled as hard spherocylinders with a diameter $d_{0}$ and length $l$ between hemispherical endcaps. To simulate growth, the length of each cell increased linearly in time until it exceeded the division length $l_{d}$, at which point the cell divided into two cells with lengths $l_{min}=(l_{d}-1)/2$. To prevent unrealistically synchronized cell division, growth rates were selected randomly in the range $[g_{0}/2,3g_{0}/2]$, where $g_{0}$ is the average cell growth rate.

Cells interacted via Hertzian contact forces, with the force on cell $i$ due to cell $j$ acting at the point of contact and calculated as follows:
\begin{equation}
\vec{F_{ij}}=Yd_{0}^{1/2}h_{ij}^{3/2}\vec{N_{ij}}
\end{equation}
where \(Y\) is proportional to the Young's modulus of a cell, \(h_{ij}\) is the overlap distance between the two cell bodies, and \(\vec{N_{ij}}\) is the vector normal to cell \(j\) at the point of contact \cite{orozcofuentes,you1,hertz}.

For force calculations, the cell backbone was defined as the line segment of length $l$ connecting the centers of the two endcaps. The overlap distance between two cells was then calculated based on the closest distance between their backbones, $h_{ij}=r_{ij}-d_{0}$, where $r_{ij}$ is the length of the shortest line segment between backbones. The point of contact was similarly defined as the midpoint of the segment $r_{ij}$.
In addition to contact forces, to prevent perfect alignment of cells, each cell was subject to a noise force at each time step with random direction and random magnitude in the range $[0,\eta_{0}]$.

Changes in cell position and orientation were modeled in the overdamped limit to reflect the assumption that cells were immersed in fluid at low Reynolds number. The equations of motion were as follows:
\begin{equation} 
\frac{d\vec{x}}{dt}=\frac{1}{l\zeta}\vec{F}
\end{equation}
\begin{equation} 
\frac{d\alpha}{dt}=\frac{12}{l^{3}\zeta}\tau
\end{equation}
where \(\vec{F}\) is the total force on the cell, \(\tau\) is the total torque, and \(\zeta\) is a drag per length originating from Stokes drag on the cells \cite{wolgemuth}. In the torque equation, $\alpha$ is the cell's orientation angle in the plane perpendicular to the net torque. These equations were numerically integrated using the explicit Euler method.

\subsection{Surface Confinement}

Cells were attached to a 2D curved surface at their centroid, with their orientation tangent to the surface at the point of contact. Based on the model system of a liquid-liquid interface, out-of-plane forces and torques were assumed to be balanced by surface tension forces. The system thus acts as if all forces are projected onto the tangent plane at the point of cell-surface contact. Similarly, torques were projected onto the unit vector normal to the surface at the contact point.

Because the surface confinement was curved and convex, the endpoints of a cell's backbone lifted slightly off the surface when it was attached and tangent to the surface at its center. Surface parameters and cell lengths were chosen so that this lift distance did not exceed the cell radius $d_{0}/2$, ensuring that cells always intersected the surface and associated errors were minimized. The actual maximum lift distance was 0.32 \textmu m.
The application of forces at each time step resulted in slight off-surface movement of cells due to the surface's curvature. To rectify this, after a cell's position and orientation were updated, its position was projected back onto the nearest point of the surface. Its orientation vector was similarly projected onto the tangent plane at that point.

Coordinates on the spherical surface were defined such that $\theta$ represents the vertical angle from the z-axis while $\phi$ represents the angle in the $x$-$y$ plane from the $x$-axis ($x=R\cos{\phi}\sin{\theta}$, $y=R\sin{\phi}\sin{\theta}$, $z=R\cos{\theta}$). The unit vectors $\hat{\theta}$ and $\hat{\phi}$ were defined similarly.

\subsection{Initial Conditions and Simulation Parameters}

Simulations were initialized with a single cell at the north pole ($\theta=0$). This simulation setup was repeated 48 times with different choices of random seed. Simulation parameters were chosen to be comparable to previous simulations of growing rods \cite{you1, langeslay1}
\begin{center}
\begin{tabular}{ |c|c| } 
 \hline
 $d_{0}$ & 1 \textmu m \\ 
  \hline
 $l_{d}$ & 6 \textmu m \\ 
  \hline
 $g_{0}$ & 3 \textmu m/h \\ 
 \hline
  $Y$ & 4 MPa \\ 
 \hline
  $\zeta$ & 200 Pa h \\ 
 \hline
\end{tabular}
\end{center}
Numerical integration was performed with a time step of 0.072s. The endpoint of the simulations was determined via the cell packing fraction $\rho$, calculated as follows:
\begin{equation}
\rho=\frac{1}{A}\sum_{i}{a_{i}}
\end{equation}
where $a_{i}$ is the lengthwise cross-sectional area of cell $i$ and $A$ is the total surface area. Simulations were terminated when $\rho$ was greater than 1.3, but the majority of analysis was conducted at $\rho=1$.

\section{Appendices}

\subsection{Appendix I: Derivation of velocity profiles for incompressible colonies}

To derive the velocity profile of cells within an expanding colony, two basic assumptions are made. First, the colony has the same symmetry as the confining geometry. Second, the density of the colony is constant in space and time, so that its area grows at the same exponential rate as the number of cells.

For the unconfined case, the symmetry assumption means that any circular region of the colony centered on the origin must be expanding evenly in all directions. The radial velocity $\dot{r}$ of cells at the edge of this circular region can be related to the rate of change of its area:
\begin{equation}
\frac{dr}{dt}=\frac{dA}{dt} \left( \frac{dA}{dr} \right) ^{-1}=(k_{g}A)(2\pi r)^{-1}=\frac{k_{g}\pi r^{2}}{2\pi r}=\frac{k_{g}}{2}r
\end{equation}
This velocity profile has previously been theoretically derived and experimentally measured in growing unconfined colonies \cite{dellarciprete}.

For the channel confined case with outlet boundaries at $x=\pm x_{o}$ and periodic boundaries at $y=\pm y_{o}$, the system is instead symmetric under translation in the $y$-direction. Therefore, instead of the circular region in the unconfined case, we choose a rectangular region centered at the origin extending to the periodic boundaries in the $y$-direction and to $\pm x$ in the $x$-direction. This rectangle must be expanding evenly along the $x$-axis. Using the same method, this gives
\begin{equation}
\frac{dx}{dt}=(k_{g}A)(4y_{o})^{-1}=\frac{k_{g}4y_{o}x}{4y_{o}}=k_{g}x
\end{equation}
This matches previous analyses of channel confined monolayers \cite{volfson, isensee}.

For the inward growing case bounded by a circular wall at radius $R$, we choose an annular region with its outer edge at $R$ and its inner edge at $r$, with area $A=\pi(R^2-r^2)$. This region must be expanding evenly in the $-r$-direction. This gives
\begin{equation}
\frac{dr}{dt}=(k_{g}A)(-2\pi r)^{-1}=\frac{k_{g}\pi(R^2-r^2)}{-2\pi r}=-k_{g}(R^{2}-r^{2})/2r
\end{equation}
This velocity profile has previously been theoretically derived and experimentally measured in inward-growing colonies, using a critical radius within a colony expanding both inward and outward in place of the bounding wall at $R$ \cite{basaran}.

\begin{figure*}
\centering
	\includegraphics[width=8.5cm]{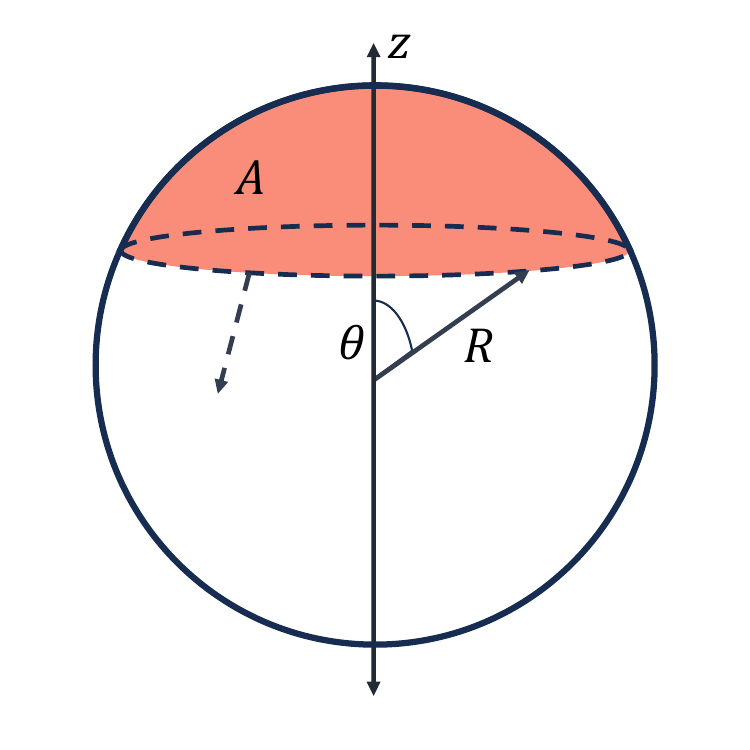}
\caption{
Schematic showing a spherical cap-shaped monolayer on a spherical surface, represented by the orange-shaded region of the sphere. The dotted boundary represents the expanding colony edge, dotted arrow represents the direction of expansion.
}
	\label{fig:figsix}
\end{figure*}

For the case of cells growing on a spherical surface of radius $R$, we choose a spherical cap centered on the initial cell at $\theta=0$ (Fig. 6). With the outer edge of the cap at $\theta$, the surface area of this region is $A=2\pi R^{2} (1-\cos{\theta})$. The cap must be evenly expanding in the $\theta$ direction, so:
\begin{equation}
\frac{d\theta}{dt}=(k_{g}A)(2\pi R^{2} \sin{\theta})^{-1}=\frac{k_{g}2\pi R^{2} (1-\cos{\theta})}{2\pi R^{2} \sin{\theta}}=k_{g}\frac{(1-\cos{\theta})}{\sin{\theta}}
\end{equation}

\subsection{Appendix II: Derivation of strain rates for incompressible colonies}

The strain component in the direction of colony expansion can be directly derived from the velocity profile:
\begin{equation}
E_{rr}=\frac{d\dot{r}}{dr}
\end{equation}
where $r$ is replaced by the applicable coordinate in the direction of expansion ($x$ in the channel case, $\theta$ in the spherical case).

The strain component in the direction perpendicular to colony expansion was derived using the perimeter $P$ of the expanding regions utilized in Appendix I. Using the definition that strain rate is equal to the rate of change in a dimension divided by the original dimension, we have
\begin{equation}
E_{\theta \theta}=\frac{1}{P}\frac{dP}{dt}
\end{equation}
where again $\theta$ is replaced with the applicable coordinate ($y$ in the channel case, $\phi$ in the spherical case).

In the unconfined case, the perimeter is $P=2\pi r$:
\begin{equation}
E_{\theta \theta}=\frac{1}{2\pi r}2\pi \dot{r}=k_{g}/2
\end{equation}
In the channel confined case, the perimeter is $P=4y_{o}$:
\begin{equation}
E_{yy}=\frac{1}{4y_{o}}(0)=0
\end{equation}
In the inward growing case, the perimeter is $P=2\pi r$:
\begin{equation}
E_{\theta \theta}=\frac{1}{2\pi r}2\pi \dot{r}=-k_{g}(R^{2}-r^{2})/2r^{2}
\end{equation}
In the spherical case, the perimeter is $P=2\pi R\sin{\theta}$:
\begin{equation}
E_{\phi \phi}=\frac{1}{2\pi R\sin{\theta}}2\pi R\dot{\theta}\cos{\theta} =\dot{\theta}/\tan{\theta}
\end{equation}

\subsection{Appendix III: Derivation of equation \ref{eqn:b} relating orientational order to strain rate}
If cells grow from $l_{min}=(l_{d}-1)/2$ to $l_{d}$ at rate $g_{0}$ before dividing, their doubling time $\tau_{d}$ is
\begin{equation}
\tau_{d}=(l_{d}-(l_{d}-1)/2)/g_{0}=\frac{l_{d}+1}{2g_{0}}
\end{equation}
leading to an exponential growth rate
\begin{equation}
k_{g}=\ln{2}/\tau_{d}=\frac{2\ln{2}g_{0}}{l_{d}+1}
\end{equation}

Meanwhile, the term $\langle 1/(l+1) \rangle$ in eqn. \ref{eqn:a} can be obtained by integrating over the total range of cell lengths:
\begin{equation}
\langle 1/(l+1) \rangle = \frac{1}{l_{d}-l_{min}} \int_{l_{min}}^{l_{d}}\frac{1}{l+1}\,dl=\frac{2}{l_{d}+1}\ln{\frac{l_{d}+1}{l_{min}+1}}=\frac{2}{l_{d}+1}\ln{2}
\end{equation}
Combining these, we have $g_{0}\langle 1/(l+1) \rangle=k_{g}$, which is substituted into eqn. \ref{eqn:a} to obtain eqn. \ref{eqn:b}.

\subsection{Appendix IV: Accounting for time-varying cell density}

If we have a time-variable cell density $\rho(t)=N/A$ for cell number $N$ and colony area $A$, then eqn. \ref{eqn:growth} no longer holds. Instead of assuming exponential growth of $A$, we must instead assume exponential growth of $N$:
\begin{equation}
\frac{dN}{dt}=k_{g}N
\end{equation}
The differential equation for $A$ is then given as follows:
\begin{equation}
\frac{dA}{dt}=\frac{1}{\rho}\frac{dN}{dt}-\frac{N}{\rho^{2}}\frac{d\rho}{dt}=\frac{k_{g}N}{\rho}-\frac{A}{\rho}\frac{d\rho}{dt}=(k_{g}-\frac{1}{\rho}\frac{d\rho}{dt})A
\end{equation}
For purposes of calculating the velocity profile and strain rates, this results in replacing the growth constant $k_{g}$ with an effective growth constant $k_{g}-\dot{\rho}/\rho$. Note that when $\dot{\rho}=0$, the original incompressible forms are recovered.

The density strain rate $\textbf{D}$ is simply the strain rate due to a uniform compression or expansion. If we hold the cell number constant, then we have
\begin{equation}
\frac{dA}{dt}=-\frac{N}{\rho^{2}}\frac{d\rho}{dt}=-A\frac{\dot{\rho}}{\rho}
\end{equation}
For an arbitrary square area $A=x^{2}$, the rate of change of a side length $x$ is as follows:
\begin{equation}
\frac{dx}{dt}=\frac{dA}{dt} \left( \frac{dA}{dx} \right) ^{-1}=-A\frac{\dot{\rho}}{\rho} \frac{1}{2x}=-\frac{x}{2}\frac{\dot{\rho}}{\rho}
\end{equation}
The density strain rate component $D_{xx}$ is then 
\begin{equation}
D_{xx}=\frac{1}{x}\frac{dx}{dt}=-\frac{1}{2}\frac{\dot{\rho}}{\rho}
\end{equation}
The same logic holds for $D_{yy}$, so the full strain rate tensor associated with this compression is
\begin{equation}
\textbf{D}=-\frac{\dot{\rho}}{2\rho}\textbf{I}
\end{equation}
\section{Author Contributions}

\textbf{Blake Langeslay:} Conceptualization, Methodology, Software, Investigation, Formal analysis, Visualization, Writing - original draft, Writing - review and editing. \textbf{Gabriel Juarez:} Conceptualization, Methodology, Supervision, Writing - original draft, Writing - review and editing.

\section{Conflicts of Interest}

There are no conflicts of interest to declare.

\section{Acknowledgements}

This work used Bridges-2 at Pittsburgh Supercomputing Center through allocation phy210132 from the Advanced Cyberinfrastructure Coordination Ecosystem: Services \& Support (ACCESS) program, which is supported by National Science Foundation grants \#2138259, \#2138286, \#2138307, \#2137603, and \#2138296 \cite{bridges2, access}.


\bibliography{bibliography}
\bibliographystyle{rsc}

\end{document}